\documentclass[twocolumn,superscriptaddress,preprintnumbers,aps,amsmath,amssymb]{revtex4}

\usepackage[T1]{fontenc}
\usepackage[latin1]{inputenc}
\usepackage{graphicx,color}
\usepackage{rotating}
\usepackage{bm}        
\usepackage{amssymb}   
\usepackage{amsmath} 
\usepackage{bm}
\usepackage{CJK}
\usepackage{ulem}
\usepackage[usenames,dvipsnames]{xcolor}

\def\jcp#1#2#3{J.~Chem.~Phys.~{\bf #1},\ #2\ (#3)}

\def\pra#1#2#3{Phys.~Rev.~A~{\bf #1},\ #2\ (#3)}

\def\k1{k_1}
\def\k2{k_2}
\def\q1{q_1}
\def\q2{q_2}

\def\({\left (}
\def\){\right )}
\def\[{\left [}
\def\]{\right ]}

\newcommand{\beq}{\begin{equation}}
\newcommand{\eeq}{\end{equation}}

\newcommand{\threejm}[6]{ \left(\begin{array}{ccc} #1 & #3 & #5\\
                                              #2 & #4 & #6
                                \end{array}
                          \right)}

\begin{document}
\date{\today}
\flushbottom \draft
\title{Tuning bimolecular chemical reactions by electric fields 
} 

\author{Timur V. Tscherbul}
\affiliation{Chemical Physics Theory Group, Department of Chemistry, and Center for Quantum Information and Quantum Control, University of Toronto, Toronto, Ontario, M5S 3H6, Canada}
\affiliation{Department of Chemistry, University of British Columbia, Vancouver, British Columbia, V6T 1Z1, Canada}
\author{Roman V. Krems}
\affiliation{Department of Chemistry, University of British Columbia, Vancouver, British Columbia, V6T 1Z1, Canada}

\begin{abstract}
{We develop a theoretical method for solving the quantum mechanical reactive scattering problem in the presence of external fields based on a hyperspherical coordinate description of the reaction complex combined with the total angular momentum representation for collisions in external fields.  The method allows us to obtain converged results for the chemical reaction LiF + H $\to$ Li + HF in an electric field. Our calculations demonstrate that, by inducing couplings between states of different total angular momenta, electric fields with magnitudes $<$150~kV/cm give rise to resonant scattering and a significant modification of the total reaction probabilities, product state distributions and the branching ratios for reactive vs inelastic scattering.}

\end{abstract}

\maketitle
\clearpage
\newpage

Tuning microscopic chemical reactions with external fields has long been an ultimate goal in chemical reaction dynamics \cite{Zare}.  
This goal stimulated the development of  quantum control schemes \cite{PaulBook,RiceZhao}, which have been applied with spectacular results to unimolecular reactions. Attaining control over bimolecular reactions in a gas has proven to be a much bigger challenge due to the randomness of the rotational and translational motion of the reactants \cite{Hershbach,RomanPCCP}. This randomness can be reduced by cooling molecules to low temperatures \cite{RomanPCCP,review-njp}, enabling the detection of quantum resonance effects in cold reactions \cite{Narevicius1,Narevicius2}. Recent experiments \cite{KRb1,KRb2,JulienneCR} demonstrated that chemical reactions in an ultracold gas of KRb molecules can be effectively suppressed by applying an electric field. While demonstrating that the randomness of the molecular motion can be harnessed, the control mechanism in Refs. \cite{KRb1,KRb2}  amounts to switching off reactive collisions by tunable long-range barriers,
 which prevent the reactants from approaching close enough to undergo chemical transformations.


In general, for chemical reactions to occur,  molecules must approach each other at close range, where the interactions induced by external fields (typically $\sim$1~K in magnitude) must compete with strong intermolecular interactions (often $>$1000~K) at short  separations between the reactants. Since the external field-induced couplings are so small compared to intermolecular interactions, it is not clear if external fields can be used to steer chemical reactions. For example, the effects of external fields on the product state distributions and branching ratios for different reaction channels remain completely unknown. While the rates of low-temperature chemical reactions can be sensitive to scattering resonances \cite{Hershbach,Narevicius1,Narevicius2}, it is not known if the resonances capable of affecting the outcome of a chemical reaction can be induced by electric or magnetic fields with feasible strengths.

These questions stimulated the mounting number of experiments  on chemical reaction dynamics in external fields \cite{JunARPC}. Several quantum threshold models  \cite{JohnBohn} and quantum defect theories  \cite{PaulJulienne,BoGao} were proposed to describe the observations. While these models provide valuable insight into  the effect of long-range interactions on ultracold reactions, with a single exception \cite{Hazra} they do not describe the reaction dynamics at short range and thus can be applied to model only the averaged quantities such as the total reaction rates.  The detailed dynamics of chemical reactions is most accurately encoded in the state-to-state scattering $S$-matrices, which can be obtained by quantum reactive scattering calculations.
However, even in the absence of external fields, the quantum reactive scattering problem is challenging due to the presence of multiple reaction arrangements  and the computational expense due to a large number of rovibrational states involved \cite{ABC,PP87,Schatz,jcp08}. The presence of external fields further complicates the problem, making it necessary to consider the coupling between states with different total angular momenta of the reaction complex. As a consequence, detailed microscopic understanding of how external fields influence the reaction mechanisms is still missing.

In this Letter, we report the first numerically exact quantum scattering calculation on a chemical reaction in an external field. Using a newly developed theoretical approach based on  hyperspherical coordinates \cite{PP87,Schatz} combined with the total angular momentum representation for collisions in external fields \cite{jcp10,pra12}, we  show that the total cross sections and the nascent product state distribution of an atom-diatom  reaction (LiF + H $\rightarrow$ Li + HF) at low collision energies can be effectively controlled by laboratory-realizable DC electric fields via tunable reactive scattering resonances. 
This work suggests that a wide range of experimentally relevant problems previously considered intractable are now amenable to theoretical study, including the effects on low-temperature chemical dynamics of reactants' spin polarization  \cite{JW}, magnetic Feshbach resonances and deviations from universality \cite{Ketterle}, and field-controlled near-resonant energy transfer \cite{jcp08,BohnNRET}.

We begin by outlining our quantum reactive scattering approach.  For the three-atom reaction considered here, there are two reaction arrangements, Li + HF and H~+~LiF that need to be considered simultaneously \cite{note1}. To do this, we use the Fock-Delves (FD) hyperspherical coordinates. Expressed in these coordinates, the Hamiltonian of the atom-molecule reaction complex in the presence of an external field is \cite{PP87,Schatz,jcp08}
\begin{equation}\label{H}
\hat{H} = \frac{-1}{2\mu \rho^5}\frac{\partial}{\partial \rho}\rho^5 \frac{\partial}{\partial\rho} + \frac{ (\hat{\bm{J}}-\hat{\bm{\jmath}}_{\alpha})^2 }{2\mu \rho^2\cos^2\theta_\alpha} + V(\rho,\theta_\alpha,\gamma_\alpha) + \hat{H}_{\text{mol},\alpha}
\end{equation}
where $\rho=(R_\alpha^2+r_\alpha^2)^{1/2}$ is the hyperradius, $\theta_\alpha$ and $\gamma_\alpha$  are the hyperangles defined by $\tan\theta_\alpha={r_\alpha}/{R_\alpha}$, and $\cos \gamma_\alpha=(\bm{R}_\alpha\cdot \bm{r}_\alpha)/(R_\alpha r_\alpha)$, and $\bm{R}_\alpha$ and $\bm{r}_\alpha$ are mass-scaled Jacobi vectors in arrangement $\alpha = 1,2,3$ \cite{PP87}.

In Eq. (\ref{H}), $\hat{\bm{J}}$ is the total angular momentum of the reaction complex and $\hat{\bm{\jmath}}_\alpha$ is the rotational angular momentum of the diatomic molecule in arrangement $\alpha$. 
The interaction of the reactants and products with the external field is included in the last term of Eq. (\ref{H}). For reactions in a DC electric field, this term is \cite{jcp08}
\begin{multline}\label{Hmol}
\hat{H}_{\text{mol},\alpha}= \frac{-1}{2\mu \rho^2\sin^22\theta_\alpha}\frac{\partial}{\partial \theta_\alpha}\sin^22\theta_\alpha \frac{\partial}{\partial\theta_\alpha} \\+ \frac{ \hat{\bm{\jmath}}_{\alpha}^2 }{2\mu \rho^2\sin^2\theta_\alpha} + V_\alpha (\rho,\theta_\alpha) - \bm{d}_\alpha(\rho,\theta_\alpha)\cdot \bm{E},
\end{multline}
where $\bm{d}_\alpha$ is the electric dipole moment of the diatomic molecule in arrangement $\alpha$ and $\bm{E}$ is the electric field vector, which defines a space-fixed (SF) quantization axis. 
The wavefunction of the reaction complex is expanded in hyperspherical adiabatic surface functions 
\begin{align}\label{BF}
\Psi = {\rho}^{-5/2} \sum_i F_i(\rho) \Phi_i(\rho;\Omega),
\end{align} 
where $\Phi_i(\Omega)$ are obtained by solving the adiabatic eigenvalue problem $\hat{H}_\text{ad}\Phi_i(\Omega;\rho)=\epsilon_i(\rho)\Phi_i(\Omega;\rho)$,  $\epsilon_i(\rho)$ are the adiabatic hyperspherical energies, and $\hat{H}_\text{ad}$ is the adiabatic surface Hamiltonian obtained by subtracting the hyperradial kinetic energy from the full Hamiltonian in Eq. (\ref{H})  \cite{PP87,Schatz,jcp08}. To solve the eigenvalue problem, we expand the surface functions as \cite{Schatz,jcp10,pra12} 
\begin{equation}\label{adiabaticExp}
\Phi_i(\rho;\Omega) = \sum_{\alpha,v,j, J,k,\eta} W_{\alpha vjJk\eta,i}  |\alpha v j Jk \eta\rangle
\end{equation}
where $|\alpha vjJk\eta\rangle = |JMk\eta\rangle {2 \chi_{\alpha vj}(\theta_\alpha;\rho)}/({\sin 2\theta_\alpha})$ 
and  $\chi_{\alpha vj}(\theta_\alpha;\rho)$ are the primitive FD basis functions, which diagonalize the Hamiltonian in Eq. (\ref{Hmol}) at {\it zero field} \cite{jcp08}.
The states $|JMk\eta\rangle$ are the angular basis functions,
\begin{equation}\label{symmetric-top}  
|JMk\eta\rangle = N_{k}\left[  |JMk\rangle |jk\rangle + \eta(-1)^{J} |JM-k\rangle |j-k\rangle  \right],
\end{equation}
composed of the spherical harmonics $|jk\rangle=\sqrt{2\pi}Y_{jk}(\theta_\alpha,0)$ and the symmetric top eigenfunctions $|JMk\rangle$, where $\eta$ is the inversion parity,  $M$ and $k$  are the projections of $J$ on the SF and body-fixed quantization axes, respectively \cite{PP87},  and $N_k=[2(1+\delta_{k0})]^{-1/2}$. 
  The basis (\ref{adiabaticExp}) is key to the efficiency of the method we propose here. 
 In an external field, $J$ and $\eta$ are not conserved but the matrix of the field-induced interaction in the basis (\ref{adiabaticExp}) is tridiagonal in $J$ and thus only a limited number of $J$-states is generally required for a fully converged calculation \cite{pra12}. This offers a great computational advantage over the previously proposed approach  \cite{jcp08}, which disregards the total angular momentum of the reaction complex.  All calculations are performed using the quantum reactive scattering program ABC \cite{ABC}, extensively modified to incorporate the effects of electric fields (see the Supplemental Material \cite{SM}).

   \begin{figure}[t]
	\centering
	\includegraphics[width=0.50\textwidth, trim = -40 0 0 0]{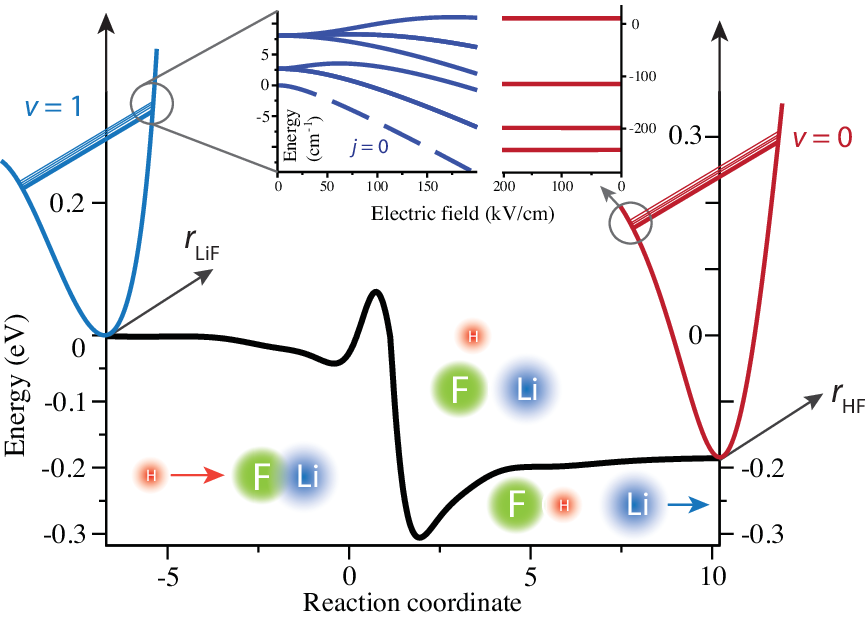}
	\renewcommand{\figurename}{Fig.}
	\caption{Schematic diagram of the LiF~+~H $\to$ HF~+~Li  chemical reaction showing  (1) the minimum energy path along the reaction coordinate $s=r_\text{LiF}-r_\text{HF}$, (2) vibrational potential energy curves of the reactants and products, and (3) the Stark structure of LiF and HF (not to scale). }\label{fig:1}
\end{figure}

We now apply this methodology to study the effects of electric fields on the chemical reaction LiF~+~H $\to$ HF~+~Li.   The choice of the reaction is motivated by the large permanent electric dipole moment of LiF ($d$ = 6.3~D), thus leading one to expect large electric field effects in the entrance reaction channel, but not in the outgoing channels \cite{jcp08}.  
In addition, the inverse reaction Li + HF $\rightarrow$ LiF~+~H has been the focus of numerous theoretical and experimental studies \cite{Althorpe03,LiHFpes,Bala05,Mudrich1}. An experimental  study of its low-temperature dynamics is in progress using a rotating nozzle source of HF molecules combined with a magneto-optical trap for Li atoms \cite{Mudrich1}. The LiF~+~H  reaction  can similarly be studied using a cold ensemble of H atoms  in a magnetic trap \cite{Doyle,Merkt}   combined with a slow beam of LiF  molecules \cite{BethlemStarkDecelerationHeavyMolecules,BufferGasBeam}. While such an experiment can be challenging to realize, we note that due to the low reduced mass of the reactants, the few-partial wave regime desirable for the observation of the effects discussed below can be reached with only moderate cooling of the reactants ($T\sim$~1~K). Collisions at such temperatures can be probed by the merged beam techniques \cite{Narevicius1,Narevicius2}.

To describe the atom-molecule interaction $V(\rho,\theta_\alpha,\gamma_\alpha)$ in the LiHF reaction complex, we use an accurate {\it ab initio} potential energy surface (PES)  \cite{LiHFpes} previously employed in field-free reaction rate calculations at low temperatures \cite{Bala05}.  
 Figure 1 illustrates the key features of the PES. The reaction proceeds  through a transition state that has a bent configuration and the barrier height is 518 cm$^{-1}$ relative to the bottom of the LiF potential well  \cite{LiHFpes}. The chemical reaction LiF$(v=1,j=0)$ + H $\to$ HF($v=0,j=0$) + Li is slightly exoergic ($\Delta E =0.1$~eV), and a total of 6 HF rotational states are energetically accessible at zero collision energy.

Figure 2 shows the total cross section for HF production in the chemical reaction of LiF$(v=1,j=0)$ with H as a function of  electric field for a collision energy of 0.01~cm$^{-1}$. At low temperatures, the reaction occurs by the tunneling of a heavy F atom \cite{Bala05} and hence the reaction cross section is small. An applied electric field causes modulation of the reaction cross section below 100~kV/cm.  
The most remarkable feature apparent in Fig. 2 is a pronounced resonance triplet at $E\sim 125$ kV/cm (peaks A, B, and C). The central resonance B corresponds to an electric-field-induced enhancement of chemical reactivity by a factor of 42. Resonances A and C have the widths of 0.10 and 0.18~kV/cm, while resonance B is at least 10 times narrower ($\Gamma \le 0.02$~kV/cm). 
To investigate the origin of these resonances, we computed the electric field dependence of the van der Waals (vdW) bound states in the   entrance reaction channel H$\cdots$LiF. We confirmed that (1) the resonances can be assigned to the bound states of the H$\cdots$LiF vdW complex, and (2) the resonances disappear if exit-channel rovibrational states are omitted from the basis set. The resonances shown in Fig. 2 are thus similar to the vdW resonances \cite{Bala05,O-HCl,FH2,Skodje} that decay via a remarkable ``pre-reaction'' mechanism involving tunnelling through the reaction barrier, even though the resonance wavefunction is localized in the entrance reaction channel \cite{Bala05}. Although the resonances acquire finite width due to  coupling to the exit reaction channel, they are sensitive to the electric field precisely because they are located in the entrance reaction channel, where the reactive system is significantly more polar.

\begin{figure}[t]
	\centering
	\includegraphics[width=0.45\textwidth, trim = 30 20 0 0]{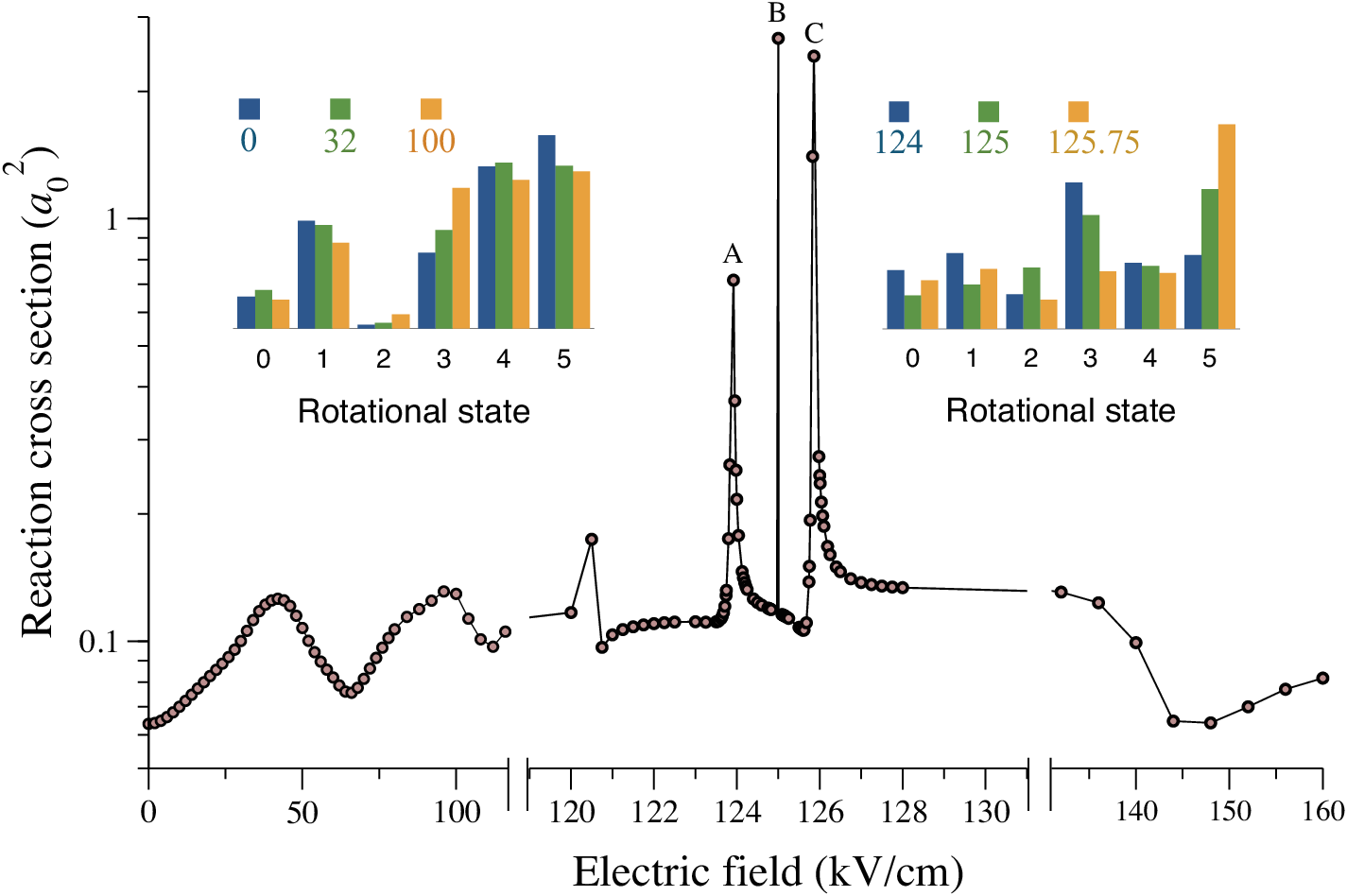}
	\renewcommand{\figurename}{Fig.}
\caption{Electric field dependence of the total cross section for the reaction LiF + H $\rightarrow$ HF + Li. The insets show the nascent rotational state distributions of HF molecules produced in the reaction as a function of the final rotational state $j'$ at electric field strengths of $0$, 32 and 100 kV/cm (left) and $124$, $125$ and $125.75$ kV/cm (right). Note the dramatic change in the shape of the distribution near the resonance electric field (right inset). All calculations were performed in the Wigner $s$-wave regime ($E_C=0.01$ cm$^{-1}$), where no resonances are present in the reaction cross sections as a function of $E_C$ \cite{Bala05}.}\label{fig:1}
\end{figure}

We next consider another important observable property of a chemical reaction, the nascent product state distribution $
\sigma_{\alpha vj \to \alpha' v'j'} / \sum_{\alpha'v'j'}\sigma_{\alpha vj \to \alpha' v'j'},
$
where $\sigma_{\alpha vj \to \alpha' v'j'}$ is the cross section for the ${\alpha vj \to \alpha' v'j'}$ reaction process. This distribution quantifies the amount of internal energy with which the reaction products form. 
Fig. 2 shows that low-to-moderate electric fields modify the rotational distributions of HF by changing the relative populations of $j' = 3$ and $j' = 5$.
  As shown below, this effect occurs due to the emergence of new chemical reaction pathways forbidden at zero fields by total angular momentum conservation.

 At $E\sim 125-127$ kV/cm corresponding to the field-induced resonances A, B, and C, the shape of the nascent product state distribution changes dramatically.  Away from the resonances, we observe a ``hot'' HF product distribution that peaks at $j'=5$ and falls off gradually with decreasing $j'$.
On resonance A, the distribution develops a pronounced peak at $j'=3$ and behaves non-monotonically as a function of $j'$, indicating a dramatic change in the reaction mechanism across a narrow interval of electric fields. On resonance B, the HF products are formed with a more even distribution over rotational energy levels, with $j'=2-5$ all substantially populated. As the electric field is tuned across resonance C, a unimodal distribution develops centered at $j'=5$. The preferential population of high $j$-states on resonances A - C can be explained by a relatively high degree of rotational excitation ($j=4$) of the LiF fragment in the vdW complex H$\cdots$LiF that gives rise to the resonance states. A more even product state distribution on resonance B results from its longer lifetime, which allows the rotational degrees of freedom to equilibrate more efficiently.  

 

\begin{figure}[t]
	\centering
	\includegraphics[width=0.47\textwidth, trim = 20 30 0 10]{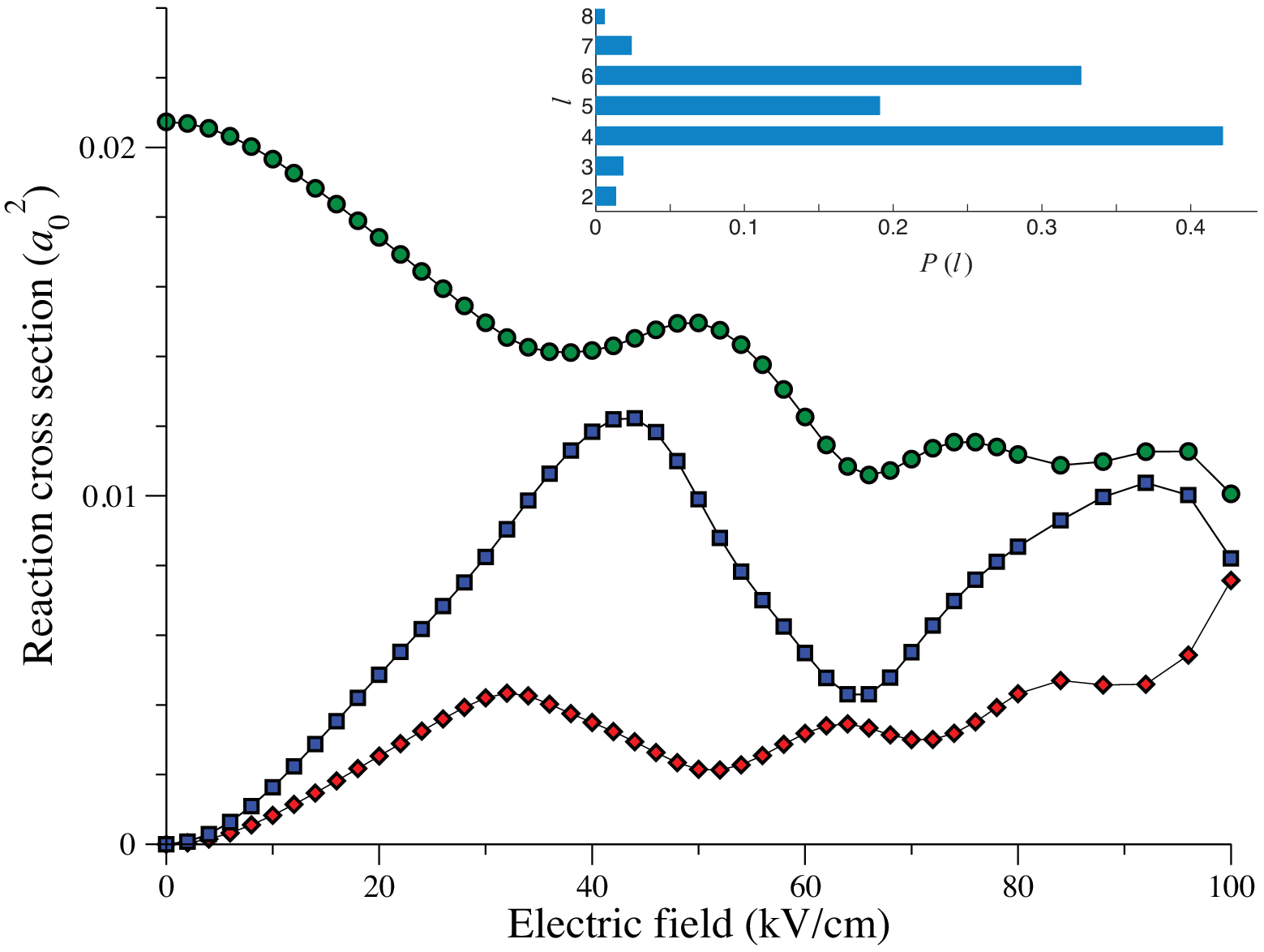}
	\renewcommand{\figurename}{Fig.}
\caption{Partial wave contributions to the cross section for the LiF($v=1,j=0$) $\to$ HF ($v'=0,j'= 5$) reactive transition as functions of an applied electric field. This transition dominates the total reaction cross section in the range of electric fields close to the resonance triplet (see Fig. 2). Circles -- $J$-conserving transition $\ell = 0 \to \ell' = 5$, squares and diamonds -- $J$-changing transitions $\ell=0 \to \ell'=4,\,6$. The inset shows the individual partial wave contribution to the $j=0 \to j'= 5$ reactive  cross section on resonance at $E=125$ kV/cm. A total of 4 $J$-states ($J=0-3$) were included in scattering calculations \cite{SM}.}\label{fig:1}
\end{figure}

In order to gain insight into the mechanism of electric field control of reaction cross sections and product state distributions, we focus on the the dominant reactive transition $j=0\to j'=5$. In Fig. 3, we plot the contributions of the different partial wave transitions $j=0,\ell=0 \to j',\ell'$ as a function of the electric field strength. Since the total angular momentum  of the collision complex $\bm{J}=\bm{j}+\bm{\ell}=\bm{j}'+\bm{\ell}'$  is conserved at zero field, and $j=\ell=0$ in the entrance reaction channel (assuming $s$-wave scattering), it follows that $j'+\ell'=0$ and hence $\ell'=j'$. Thus only the $\ell'=5$ partial wave contribution is allowed at zero field. The line with circles in Fig. 3 confirms this. An external field induces couplings between the adjacent $J$ states \cite{jcp08,pra12}. As a result, the off-diagonal, $J$-changing transitions $j=0,\ell=0 \to j', \ell' = j' \pm 1$ become allowed, 
as  illustrated in Fig.~3. While these $J$-changing transitions play a minor role at low fields, they become dominant at fields above 100 kV/cm. As shown in the inset of Fig. 3, the $J$-changing transitions $\ell=0\to \ell'=4,6$ make up more than 70\% of the reaction cross section at $E=125$ kV/cm (on resonance B). We therefore refer to resonance B as the {\it electric-field-induced} resonance.

While the electric-field-induced resonances can greatly enhance the reaction cross section, the excess vibrational energy of the LiF($v=1,j=0$) reactants can also be converted into translational energy via non-reactive collisions leading to vibrational relaxation. 
 To explore the possibility of controlling the relative efficiency of these competing pathways,  we plot in Fig.~4 the electric field dependence of the ratio of cross sections for vibrational relaxation and reactive scattering. At low fields, the branching ratio varies insignificantly, and vibrational relaxation remains as efficient as it is at zero field. Near the electric field-induced resonance, however, the branching ratio drops to 4 before raising back to 10.

The electric field dependence of the LiF$(v=0,j')$ product distribution following vibrational relaxation in LiF$(v=1,j=0)$ + H collisions is plotted in the inset of Fig.~4 as a function of $j'$. We observe strong variation of the distributions even at low electric fields. A moderate field of 40~kV/cm broadens the distribution significantly, populating higher $j'$-states. 
We attribute this effect to the field-induced hybridization of LiF rotational states in the $v=0$ manifold, which modifies the anisotropic part of the LiF-H interaction potential and changes the relative populations of final rotational states.  At the resonance B, the rotational distribution becomes extremely broad and multimodal. While transitions to high-$j$ states are suppressed at low-to-moderate electric fields, they become allowed at $E=125$~kV/cm, signalling a profound change in the mechanism of rovibrational energy transfer near electric field-induced scattering resonances. This mechanism is different from that explored in previous work on near-resonant energy transfer in cold collisions \cite{BalaVR} as the energy gaps between the rovibrational levels of the reactants and products remain large ($>$20~cm$^{-1}$) in the range of electric fields explored in this work.

\begin{figure}[t]
	\centering
	\includegraphics[width=0.48\textwidth, trim = 20 20 0 -10]{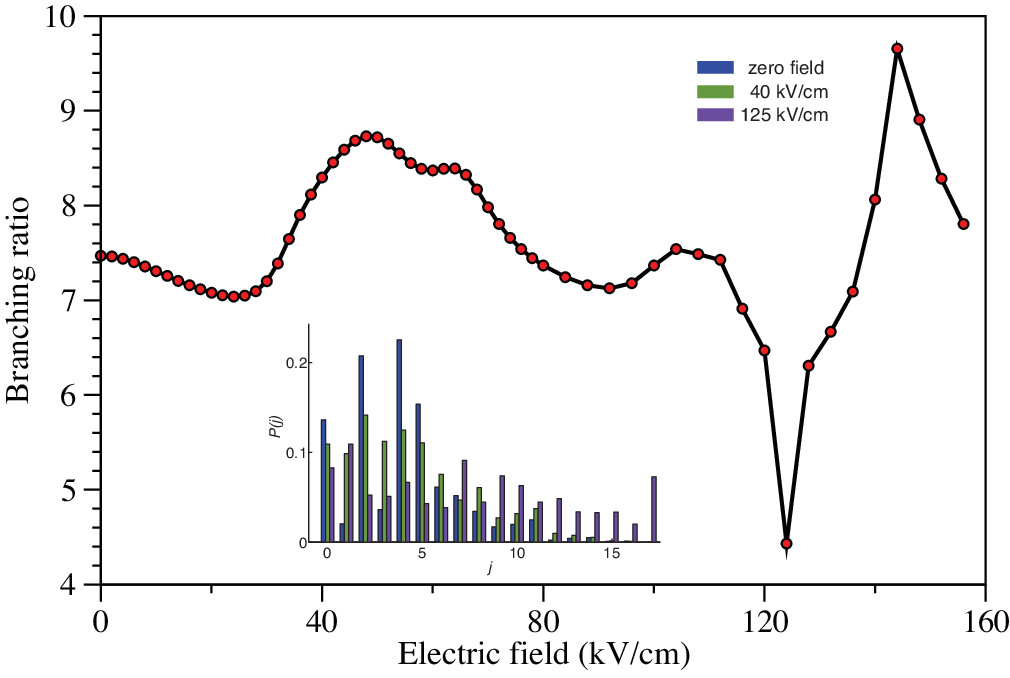}
	\renewcommand{\figurename}{Fig.}
	\caption{The branching ratio  for inelastic to reactive cross sections as a function of electric field. The inset shows rotational product state distributions for vibrational relaxation in non-reactive LiF($v=1,j=0$) + H collisions.   }\label{fig:1}
\end{figure}



In conclusion, we have introduced a theoretical method for solving the quantum reactive scattering problem in the presence of an external field  based on a  hyperspherical coordinate formalism \cite{PP87,Schatz,ABC} combined with the total angular momentum representation for molecular collisions in external fields \cite{jcp10,pra12}.  The method is much more efficient than the previous rigorous approach \cite{jcp08} and makes it possible to obtain numerically converged results for a three-dimensional atom-diatom chemical reaction in a DC electric field. The efficiency can be further enhanced by transforming away the off-diagonal $J$-blocks, or by matching to quantum defect solutions \cite{Hazra}. Our methodology can be applied to any abstraction atom-diatom chemical reaction in magnetic, DC electric and off-resonant microwave and laser fields. It can also be extended to calculations on barrierless insertion chemical reactions by changing the hyperspherical part of the treatment to the Smith-Whitten coordinates \cite{SWcoordinates}. The main idea of combining the field-free reactive scattering problems formulated in the $J$-representation and including field-dependent couplings between different $J$-states would still apply.


This work was supported by NSERC of Canada. We are grateful to D. Ding for his expert assistance with high-performance computing.


\newpage

\begin{centering}
{\bf Supplemental Material for the manuscript \\ ``Tuning bimolecular chemical reactions by electric fields''}
\end{centering}



This Supplemental Material provides details for the numerical implementation of quantum reactive scattering calculations in the presence of an external electric field. In Sec. I we give the explicit expressions for the molecule-field interaction matrix elements in the parity-adapted total angular momentum basis (see main text for details) and describe how these expressions are implemented in the extended ABC code, which we developed to study the low-temperature dynamics of atom-diatom chemical reactions in the presence of electric fields.
Section II describes the procedure of solving the scattering equations and focuses on the derivation and implementation of reactive scattering boundary conditions in the presence of external fields. In order to verify the implementation of the boundary conditions, we present in Sec. IIIA test calculations of the cross sections for collision-induced rotationally inelastic transition  LiF$(v=1,j=1)$ + H $\to$ LiF($v=1,j=0$) + H in an electric field. The results of this calculation agree with independent calculations using a different scattering code \cite{pra12}. Finally,  Sec. IIIB describes convergence tests performed and justifies the choice of    convergence parameters for the computation of reaction cross sections.

\section{Detailed equations used in the computer code}

As discussed in the main text, the interaction of the reactants and products with an external electric field can be incorporated in quantum reactive scattering theory by including the molecule-field interaction term (see Eq. (2) of the main text)
\begin{equation}\label{Hef}
\hat{H}_{E} = -\bm{d}_\alpha \cdot \bm{E}
\end{equation}
This form of the molecule-field interaction is valid assuming that (1) electric field-induced coupling between the different reaction arrangements can be neglected and (2) the dipole moment of the diatomic molecule $d$ is assumed independent of $r_\alpha$. The first approximation is well-justified because the molecule-field interaction is extremely weak compared to the chemical interactions at short-range ($\rho < 7$ $a_0$) where the atoms are close together and the chemical reaction takes place. The second approximation holds for low-lying vibrational states of the diatomic molecule usually involved in reactive collisions.

The matrix elements of the molecule-field coupling (\ref{Hef}) in the primitive Fock-Delves (FD) hyperspherical basis $|\alpha v j Jk \eta \rangle$ (see main text for details)
can be evaluated by transforming the angular part of the basis to a space-fixed coordinate frame and using standard angular momentum algebra \cite{pra12}. The final result is
\begin{multline}\label{me}
\langle \alpha v j Jk\eta | \hat{H}_{E}| \alpha' v' j' J' k'\eta'\rangle =   
-d_\alpha E \langle  \chi_{\alpha vj}(\theta_\alpha;\rho)| \chi_{\alpha v'j'}(\theta_{\alpha};\rho)\rangle   \\ \times
\frac{\delta_{\alpha\alpha'} }{[(1+\delta_{k0})(1+\delta_{k'0})]^{1/2}} [(2J+1)(2J'+1)(2j+1)(2j'+1)]^{1/2} 
 \\ \times (-1)^M \delta_{\eta+\eta',0} \threejm{J}{M}{1}{0}{J'}{-M} \threejm{j}{0}{1}{0}{j'}{0} \\ \times
\biggl{[} \threejm{J}{k}{1}{k'-k}{J'}{-k'} \threejm{j}{-k}{1}{k-k'}{j'}{k'}  \\ + \eta' (-)^{J'} \threejm{J}{k}{1}{-k'-k}{J'}{k'} \threejm{j}{-k}{1}{k+k'}{j'}{-k'}   \biggr{]}
\end{multline}
The molecule-field interaction is thus non-diagonal in $J$ and $\eta$, due to the breaking of the inversion symmetry  by electric fields. The matrix element in Eq. (\ref{me}) is a product of the angular part represented by sums of products of 3-$j$ symbols multiplied by the hyperangular overlap of the primitive FD basis functions 
\begin{equation}\label{HApart}
\langle  \chi_{\alpha vj}(\theta_\alpha;\rho)| \chi_{\alpha v'j'}(\theta_{\alpha};\rho)\rangle
\end{equation}

 We further note that the  matrix elements (\ref{me}) vanishe unless $j=j\pm1$, $J=J\pm 1$, and $\eta'=-\eta$; thus, an external DC electric field hybridizes the adjacent rotational states of the reactants and products. The effects of field-induced orientation are particularly pronounced in the entrance reaction channel. 

The molecule-field interaction matrix elements (\ref{me}) are evaluated by two newly written subroutines $\mathtt{field\_ov\_FD}$ and $\mathtt{addElectricField}$ added to the ABC code.  $\mathtt{addElectricField}$  adds the molecule-field interaction matrix elements to the field-free part of the adiabatic surface (AS) Hamiltonian matrix in the primitive FD basis (see main text for details) to produce the total field-dependent AS Hamiltonian matrix, which is then diagonalized to yield the surface functions. The field-free AS Hamiltonian matrix is constructed by the subroutines $\mathtt{direct}$ and $\mathtt{exchng}$ in the original ABC code \cite{ABC}.  

The matrix element   (\ref{me}) is computed in two steps. First, the subroutine $\mathtt{field\_ov\_FD}$ calculates the hyperangular overlaps (\ref{HApart}) for a given $J,\eta$, $J',\eta'$, and $\alpha$, by expanding the Fock-Delves basis functions over primitive particle-in-a-box eigenfunctions.  
Secondly, the subroutine $\mathtt{addElectric field}$  evaluates the full molecule-field interaction matrix element in Eq. (\ref{me}) by multiplying the hyperangular overlap matrix element with  the 3-$j$ symbols.  The resulting molecule-field interaction matrix element is added to the field-free AS Hamiltonian.


\section{Boundary conditions for numerical calculations}

The hyperradial expansion coefficients $F_i(\rho)$ in Eq.~(5) of the main text satisfy a system of coupled second-order differential equations  parametrized by the matrix elements of the Hamiltonian [Eq. (1) of the main text]  and we use the log-derivative algorithm \cite{LDA} to integrate the equations on a grid of $\rho$ sectors. At the intersector boundary the wavefunction is transformed to the hyperangular basis of the next sector using the overlap matrix $\langle \Phi_i(\Omega;\rho_k)|\Phi_j(\Omega;\rho_{k+1})\rangle$ constructed from the basis set expansion given by Eq. (3) of the main text. After reaching the asymptotic region of large hyperradius $\rho_\text{max}$, the wavefunction of the reactive complex is projected onto the eigenfunctions of the atom-molecule system in Jacobi coordinates to extract  the reaction probabilities and cross sections as described below.


At $\rho=\rho_\text{max}$, the wave function of the reaction complex is expanded in field-dressed FD hyperangular basis functions 
 \begin{equation}\label{FDexpansion}
|\Psi^{i} \rangle_\text{FD} = \rho^{-5/2} \sum_{f} \Gamma^{i}_{f} (\rho) |f \rangle_\text{FD}
\end{equation}
where 
 \begin{equation}\label{FDbasis1}
|f\rangle_\text{FD} = \sum_{n'} C^\text{FD}_{n'f}  |n'\rangle,
\end{equation}
are the field-free (primitive) FD basis functions in a space-fixed (SF) coordinate frame given by
 \begin{equation}\label{FDbasis2}
|n'\rangle =  \left[ \frac{2 \chi_{\alpha_{n'} v_{n'}j_{n'}}(\theta_{\alpha_{n'}};\rho)}{\sin 2\theta_{\alpha_{n'}}}  \right]  \mathcal{J}^{J_{n'}M}_{j_{n'}l_{n'}} (\hat{R}_{\alpha_{n'}};\hat{r}_{\alpha_{n'}}).
\end{equation}
where $\chi_{\alpha_{n'} v_{n'}j_{n'}}(\theta_{\alpha_{n'}};\rho)$ are the primitive FD hyperspherical basis functions (see main text for details) and
 \begin{multline}\label{FDbasisBipolar}
\mathcal{J}^{JM}_{jl} (\hat{R}_{\alpha};\hat{r}_{\alpha}) = \sum_{m_j, \,m_l} (-1)^{j-l+M} (2J+1)^{1/2} \\ \times
\threejm{j}{m_j}{l}{m_l}{J}{-M} Y_{jm_j}(\hat{r}_\alpha) Y_{lm_l}(\hat{R}_\alpha)
\end{multline}
are bipolar spherical harmonics  \cite{PP87} (the $n'$ subscripts have been omitted for clarity). 
In Eq. (\ref{FDbasis1}),  $C^\text{FD}_{n'f}$ are the Stark mixing coefficients that can be obtained by diagonalizing the asymptotic Hamiltonian $\hat{H}_\text{as}$ expressed in the primitive SF basis (\ref{FDbasis2}).  
For the sake of simplicity throughout this section, we label field-dressed basis functions with Latin letters, {\it e.g.},  $n = \{ \alpha_n, v_n, \gamma_n, l_n \}$,  $i = \{ \alpha_i, v_i, \gamma_i, l_i \}$, and so on (note that the index $i$ corresponds to the incident scattering channel).  The primed indices are reserved for field-free (primitive) basis functions, {\it e.g.}, $f' = \{\alpha_{f'}, v_{f'}, j_{f'}, l_{f'},J_{f'}\}$,  $n' = \{\alpha_{n'}, v_{n'}, j_{n'}, l_{n'}, J_{n'}\}$, etc.

While the representation given by Eq. (\ref{FDexpansion}) is ideal for numerical solution of close-coupled differential calculations, it does not easily lend itself to the asymptotic analysis required to  extract the $S$-matrix elements for reactive transitions between the individual Stark  states of the reactants and products \cite{PP87}. In order to compute the $S$-matrix elements, we need to transform the wavefunction to a representation that diagonalizes the asymptotic Hamiltonian. To this end,  we use an expansion in Jacobi coordinates \cite{PP87} suitably generalized to include the modification of channel wavefunctions by external fields \cite{RomanAlex04,pra12}
\begin{equation}\label{Jexpansion}
|\Psi^{i} \rangle_\text{Jac} = \sum_{n} \frac{1}{R_{\alpha_n} r_{\alpha_n}}  F^{i}_{n} (R_{\alpha_n})  |n \rangle_\text{Jac} 
\end{equation}
where
\begin{multline}\label{Jac}
|n \rangle_\text{Jac} = |\alpha_n v_n \gamma_n l_n \rangle_\text{Jac} =  \sum_{n'} C^\text{Jac}_{n'; n} \\ \times\xi_{\alpha_{n'} v_{n'} j_{n'}} (r_{\alpha_{n'}})  \mathcal{J}^{J_{n'} M}_{j_{n'} l_{n'}} (\hat{R}_{\alpha_{n'}};\hat{r}_{\alpha_{n'}})
\end{multline}
are the field-dressed basis functions in Jacobi coordinates, $ \xi_{\alpha_{n'} v_{n'} j_{n'}} (r_{\alpha_{n'}})$ is the rovibrational eigenfunction of the molecule in arrangement $\alpha_{n'}$ characterized by the vibrational and rotational quantum numbers $v_{n'}$ and $j_{n'}$.

It follows from Eq. (\ref{FDexpansion}) and the orthogonality property of field-dressed FD basis functions (\ref{FDbasis1}) that for sufficiently large $\rho$
 \begin{equation}
 \Gamma^i_f(\rho)=\Gamma_{fi}(\rho)  = \langle f | \rho^{5/2} \Psi^{i}  \rangle_\text{FD}
 \end{equation}
To perform the coordinate transformation, we substitute  $|\Psi^{i}  \rangle_\text{Jac}$ from Eq. (\ref{Jexpansion}) for $|\Psi^{i}  \rangle_\text{FD}$ \cite{PP87} and use the orthonormality properties of bipolar spherical harmonics to obtain after some algebra
 \begin{equation}\label{Gamma6}
\Gamma_{fi} (\rho) = \sum_{n} \sum_{f',n'} C_{f' f}^\text{FD} C_{n' n}^\text{Jac}  \Lambda^{ni}_{f'n'}
\end{equation}
where
\begin{multline}\label{Lambda}
\Lambda^{ni}_{f'n'} = 
\delta_{\alpha_{f'}\alpha_{n'}} \delta_{j_f' j_n'}\delta_{l_{f'} l_{n'}}\delta_{J_{f'} J_{n'}} 
 \rho^{1/2}  \int_0^{\pi/2} 
 \chi_{\alpha_{f'}v_{f'}j_{f'}}(\theta_{\alpha_{f'}};\rho) \\ \times  F^{\alpha_i v_i \gamma_i l_i}_{\alpha_n v_n \gamma_n l_n}(R_{\alpha_{f'}}) \xi_{\alpha_{n'} v_{n'} j_{n'}} (r_{\alpha_{f'}}) d\theta_{\alpha_{f'}} .
\end{multline}
is a tensor of rank 4 that depends on the field-free (primed) as well as field-dressed (unprimed) indexes of basis functions.

In the limit of large atom-molecule separation $R_{\alpha_{n'}}$, the radial expansion coefficients $F$ take the form
\begin{equation}\label{BoundaryConditions}
F_{ni}(R_{\alpha_f'}) \to \delta_{ni} a_n(R_{\alpha_{f'}}) - b_n(R_{\alpha_{f'}}) K_{ni},
\end{equation}
where $K_{ni}$ are the $K$-matrix elements and the functions $a_n$ and $b_n$ are proportional to the Riccati-Bessel functions or modified Bessel functions of the third kind depending on whether the asymptotic scattering channel $n$ is open or closed \cite{PP87}.

Using Eq. (\ref{BoundaryConditions}) we obtain for the $\Lambda$-tensor
\begin{equation}\label{Lambda2}
\Lambda^{ni}_{f'n'} = \delta_{ni}\mathcal{A}^n_{f'n'} - \mathcal{B}^n_{f'n'}K_{ni}.
\end{equation}
where
\begin{multline}\label{A}
\mathcal{A}^{n}_{f'n'} = 
\delta_{\alpha_{f'}\alpha_{n'}} \delta_{j_{f'} j_{n'}}\delta_{l_{f'} l_{n'}}\delta_{J_{f'} J_{n'}} 
 \rho^{1/2} \\ \times  \int_0^{\pi/2} d\theta_{f'}
 \chi_{f'}(\theta_{f'};\rho)  a_{n}(R_{f'}) \xi_{n'} (r_{f'}). 
 \end{multline}
 \begin{multline}\label{B}
\mathcal{B}^{n}_{f'n'} = 
\delta_{\alpha_{f'}\alpha_{n'}} \delta_{j_{f'} j_{n'}}\delta_{l_{f'} l_{n'}}\delta_{J_{f'} J_{n'}} 
 \rho^{1/2} \\ \times  \int_0^{\pi/2} d\theta_{f'}
 \chi_{f'}(\theta_{f'};\rho)  b_{n}(R_{f'}) \xi_{n'} (r_{f'}).
\end{multline}
are tensors of rank 3.
We note that unlike Eqs. (116) and (117) of Ref. \cite{PP87}, the radial functions $a_n$ and $b_n$ in Eqs.~(\ref{A}) and (\ref{B}) are given in a field-dressed scattering basis. In particular, the wavevectors $k_n$ entering the arguments of the functions $a_n$ and $b_n$ correspond to the states of the reactants and products in the presence of an electric field, while the basis functions $f'$ and $n'$ are the field-free basis functions used in conventional quantum reactive scattering theory \cite{PP87}. As a result, the  quantities $\mathcal{A}$ and $\mathcal{B}$ in Eqs. (\ref{A}) and (\ref{B}) acquire an extra index $n$. In the limit of zero electric field there is no coupling between the rotational states of the reactants or products, so $C_{n'n}^\text{FD}=C_{n'n}^\text{Jac}=\delta_{n'n}$, and Eqs.~(\ref{A}) and (\ref{B}) reduce to Eqs. (116) and (117) of Ref. \cite{PP87}.  

Defining the matrix-tensor products
\begin{align}\label{MT}\notag
[\mathbf{C}^T \mathcal{A}^n\mathbf{C}]_{fn} &=  \sum_{f',n'} \delta_{ni} (C^\text{FD})^T_{ff'} \mathcal{A}^n_{f'n'} C^\text{Jac}_{n'n}, \\
[\mathbf{C}^T \mathcal{B}^n\mathbf{C}]_{fn} &= \sum_{f',n'}  (C^\text{FD})^T_{ff'} \mathcal{B}^n_{f'n'} C^\text{Jac}_{n'n}.
\end{align}
and using Eq. (\ref{Lambda2}), we can bring the expression (\ref{Gamma6}) to the form
\begin{equation}\label{GammaMatrix2}
{\Gamma}_{fi}(\rho) =  [\mathbf{C}^T \mathcal{A}^i\mathbf{C}]_{fi} -  \sum_n [\mathbf{C}^T  \mathcal{B}^n\mathbf{C}]_{fn} {K}_{ni} 
\end{equation}
or in matrix form
\begin{equation}\label{GammaFinal}
\mathbf{\Gamma}(\rho) =  [\mathbf{C}^T \mathcal{A}\mathbf{C}] -   [\mathbf{C}^T \mathcal{B}\mathbf{C}] \mathbf{K}
\end{equation}
We observe that the matrix-tensor products in square brackets are square $N\times N$ matrices (where $N$ is the number of channels). Because $n$ is not only a summation index  but also determines the form of $\mathcal{B}^n$ itself,  the matrix-tensor products in Eq. (\ref{GammaFinal}) are more  difficult to compute than the usual matrix-matrix products.

The integration of coupled differential equations by the log-derivative algorithm \cite{LDA} produces the log-derivative matrix at a large value of $\rho$
\begin{equation}\label{Ydef}
\mathbf{Y}  =\, \frac{d \mathbf{\Gamma(\rho)}}{d\rho} [\mathbf{\Gamma(\rho)}]^{-1}
\end{equation}
where $\mathbf{\Gamma}$ is the matrix of hyperradial coefficients $\Gamma_{fi}(\rho)$. 
In order to extract the $K$-matrix from the log-derivative matrix, we need to evaluate the wavefunction of the reaction complex and its radial derivative in Jacobi coordinates. We have already completed the first step [see Eq. (\ref{GammaFinal})]. Taking the first derivative of the wavefunction matrix (\ref{GammaFinal}) with respect to $\rho$, we obtain after a sequence of transformations  \cite{TBP}
\begin{equation}\label{matrix_dGamma}
\frac{d \mathbf{\Gamma (\rho)}}{ d\rho} = \frac{1}{2\rho} \mathbf{\Gamma}(\rho) + [\mathbf{C}^T \mathcal{G} \mathbf{C}] -  [\mathbf{C}^T \mathcal{H} \mathbf{C}]\mathbf{K},
\end{equation}
where 
\begin{multline}\label{G}
\mathcal{G}^n_{f'n'}= \delta_{\alpha_{f'}\alpha_{n'}} \delta_{j_{f'} j_{n'}}\delta_{l_{f'} l_{n'}}\delta_{J_{f'} J_{n'}} 
 \rho^{1/2}  \int_0^{\pi/2}
 \chi_{f'}(\theta_{f'};\rho) \\ \times   \Biggl{[} \frac{\partial a_n(R_{f'})}{\partial R_{f'}} \xi_{n'} (r_{f'}) \cos\theta_{f'} + a_n(R_{f'}) \frac{\partial \xi_{n'}(r_{f'})}{\partial r_{f'}} \sin\theta_{f'} \Biggr{]}  d\theta_{f'}
\end{multline}
and
\begin{multline}\label{H}
\mathcal{H}^n_{f'n'}  = \delta_{\alpha_{f'}\alpha_{n'}} \delta_{j_{f'} j_{n'}}\delta_{l_{f'} l_{n'}}\delta_{J_{f'} J_{n'}} 
 \rho^{1/2} \int_0^{\pi/2} \chi_{f'}(\theta_{f'};\rho)  \\ \times
 \Biggl{[} \frac{\partial b_n(R_{f'})}{\partial R_{f'}} \xi_{n'} (r_{f'}) \cos\theta_{f'} + b_n(R_{f'}) \frac{\partial \xi_{n'}(r_{f'})}{\partial r_{f'}} \sin\theta_{f'} \Biggr{]}  d\theta_{f'}.
\end{multline}
are rank-3 tensors similar in structure to $\mathcal{A}$ and $\mathcal{B}$. Combining Eqs. (\ref{matrix_dGamma}) and (\ref{GammaFinal}) and solving for $\mathbf{K}$, we find
\begin{multline}\label{K}
\mathbf{K} = \left\{ \left(\mathbf{Y}(\rho_m) -\frac{1}{2\rho}\mathbf{I}\right) [\mathbf{C}^T \mathcal{B}\mathbf{C}] - [\mathbf{C}^T \mathcal{H}\mathbf{C}]   \right \}^{-1}  \\ \times
 \left\{ \left(\mathbf{Y}(\rho_m) -\frac{1}{2\rho}\mathbf{I}\right) [\mathbf{C}^T \mathcal{A}\mathbf{C}] - [\mathbf{C}^T \mathcal{G}\mathbf{C}]   \right \}.
\end{multline}

In the extended version of the ABC code, the matrix-tensor products are evaluated in a Gauss-Legendre quadrature loop over $\theta_{\alpha}$ for each reaction arrangement. The tensors $\mathcal{A}$, $\mathcal{B}$, ${\mathcal{G}}$, and ${\mathcal{H}}$ are computed in two loops over $f'$ and $n'$ added inside the $\theta_\alpha$ quadrature loop. In follows from Eq. (\ref{A}) and related expressions that most of the off-diagonal tensor elements of $\mathcal{A}$, $\mathcal{B}$, $\mathcal{G}$, and $\mathcal{H}$ are zero. To exploit this sparseness, we nest the $f'$ and $n'$ loops in such a way as to ensure that summation over $n'$ and $f'$ includes only nonzero tensor elements in Eq. (\ref{MT}), thereby leading to a substantial decrease in computational effort.   Even with the sparse structure of  projection tensors taken into account,  the evaluation of Eqs. (\ref{MT}) is computationally intensive. The computational cost of constructing the transformation matrices grows nonlinearly with increasing basis set size, but remains modest for basis sets with $N\le 2500$.

The $K$-matrix is computed using Eq. (\ref{K}), and then converted to the $S$-matrix using the expression $\mathbf{S}=(\mathbf{I}+i\mathbf{K}^{oo})(\mathbf{I}-i\mathbf{K}^{oo})^{-1}$, where $\mathbf{K}^{oo}$ is the open-open block of the $K$-matrix and $\mathbf{I}$ is the unit matrix \cite{PP87}.
 The reaction cross sections are calculated from the $S$-matrix using the expression
 \begin{equation}\label{sigma}
\sigma_{\alpha v \gamma  \to \alpha'v' \gamma' } = \frac{\pi}{k^2_{\alpha v\tilde{\jmath}} }\sum_M \sum_{l,\, l'} P^M_{\alpha v \gamma l  \to \alpha'v' \gamma'l' }
\end{equation}
where 
\begin{align}\label{P}
P^M_{\alpha v \gamma l  \to \alpha'v' \gamma'l' } &= | S^M_{\alpha v \gamma  l \to \alpha'v' \gamma' l'}  |^2   \quad (\alpha\ne \alpha')
\end{align}
is the fully state-resolved reaction probability  and the index $\gamma$ runs over the Stark states of the reactants and products (note that in the zero-field limit, the index $\gamma$ can be replaced with $j$, $J$ and $\eta$, and transitions between the states with different $J$ and $\eta$ become forbidden).

\begin{figure}[t!]
	\centering
	\includegraphics[width=0.6\textwidth, trim = 90 0 0 0]{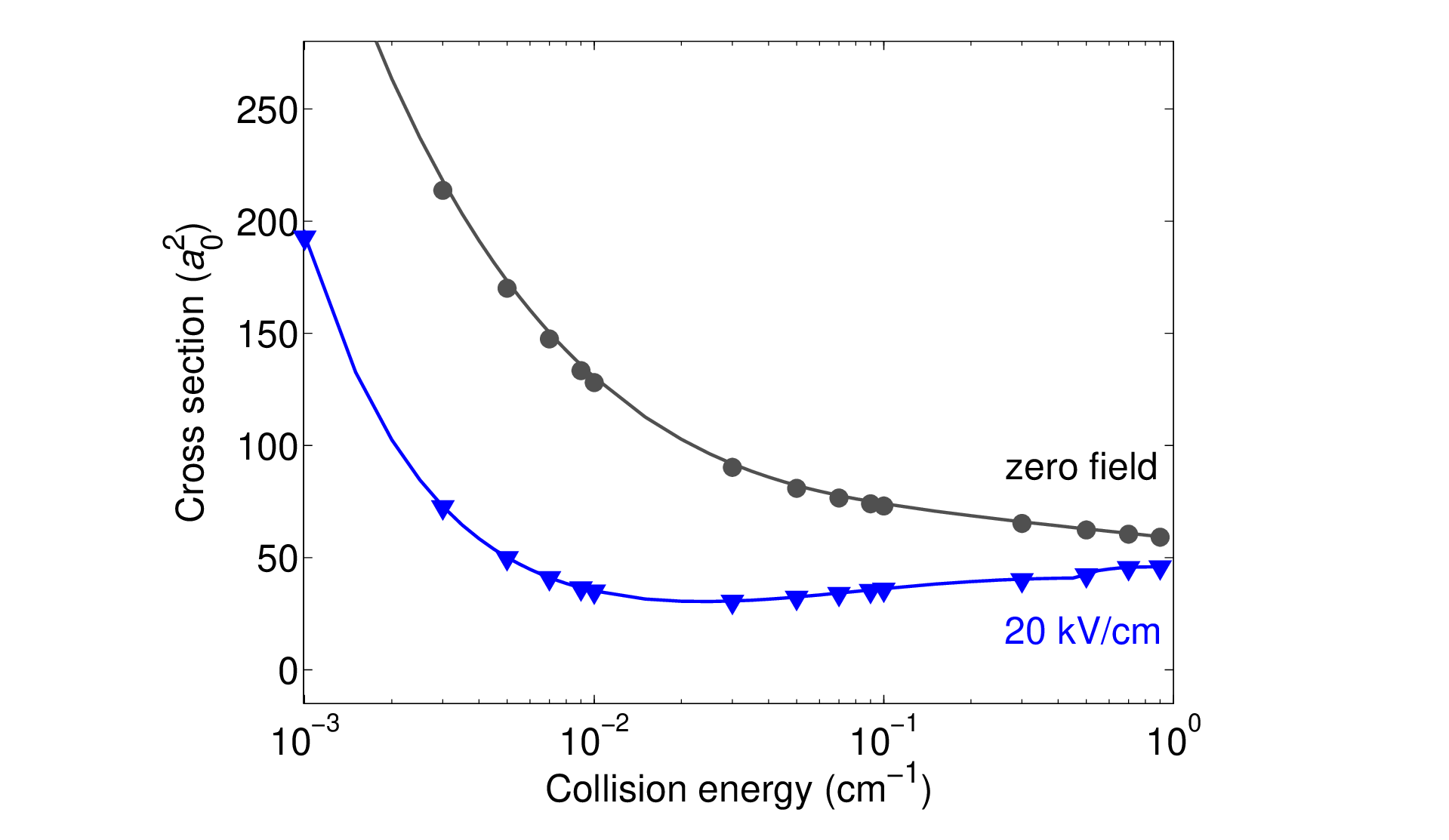}
	\renewcommand{\figurename}{Fig.}
	\caption{Cross sections for rotational relaxation in LiF($v=1,j=1$) + H $\to$ LiF($v=1,j=0$) + H collisions as functions of collision energy in the absence of external fields (top trace) and for $E= 20$ kV/cm (bottom trace). Symbols -- calculations using the extended ABC code; lines -- benchmark results obtained with the non-reactive scattering code of Ref. \cite{pra12}. }\label{fig:1}
\end{figure}

\section{Numerical tests of calculation accuracy}

\subsection{Extended ABC code tests}

In order to verify the extensive modifications made to the ABC code to incorporate the effects of electric fields, we carried out two separate series of test calculations. In order to ensure reliable performance of the code in the absence of an electric field, we calculated the reactive scattering cross sections  for LiF$(v=1,j=0)$ + H $\to$ Li + HF as a function of collision energy  using the original (unmodified) version of the ABC code \cite{ABC}. The resulting cross sections were properly summed over $J$ and compared with the cross sections computed using the extended ABC code (properly summed over $M$). The extended version of the ABC code and the original ABC code \cite{ABC} were found to produce identical results, thereby ensuring proper  implementation of the multiple-$J\eta$ hyperspherical FD basis set (see Eq. (4) of the main text).

To test the performance of the code in the presence of an electric field, we calculated the cross sections for rotational relaxation  in LiF$(v=1,j=1)$  + H $\to$ LiF($v=1,j=0$) + H using a different scattering code developed in Ref.~\cite{pra12} for non-reactive atom-molecule collisions in electric fields. Figure 5 demonstrates good agreement between the inelastic cross sections produced by the code developed in Ref.~\cite{pra12} and those obtained using the extended ABC code. Given that the code used in Ref. \cite{pra12} employs a different coordinate system to represent the scattering wavefunction, and a different expansion for the interaction potential, the agreement  strongly suggests that the molecule-field interaction Hamiltonian (Sec. I) and the boundary conditions (Sec. II) have been implemented correctly.

\subsection{Convergence tests}

 \begin{figure}[t!]
	\centering
	\includegraphics[width=0.43\textwidth, trim = 90 0 0 0]{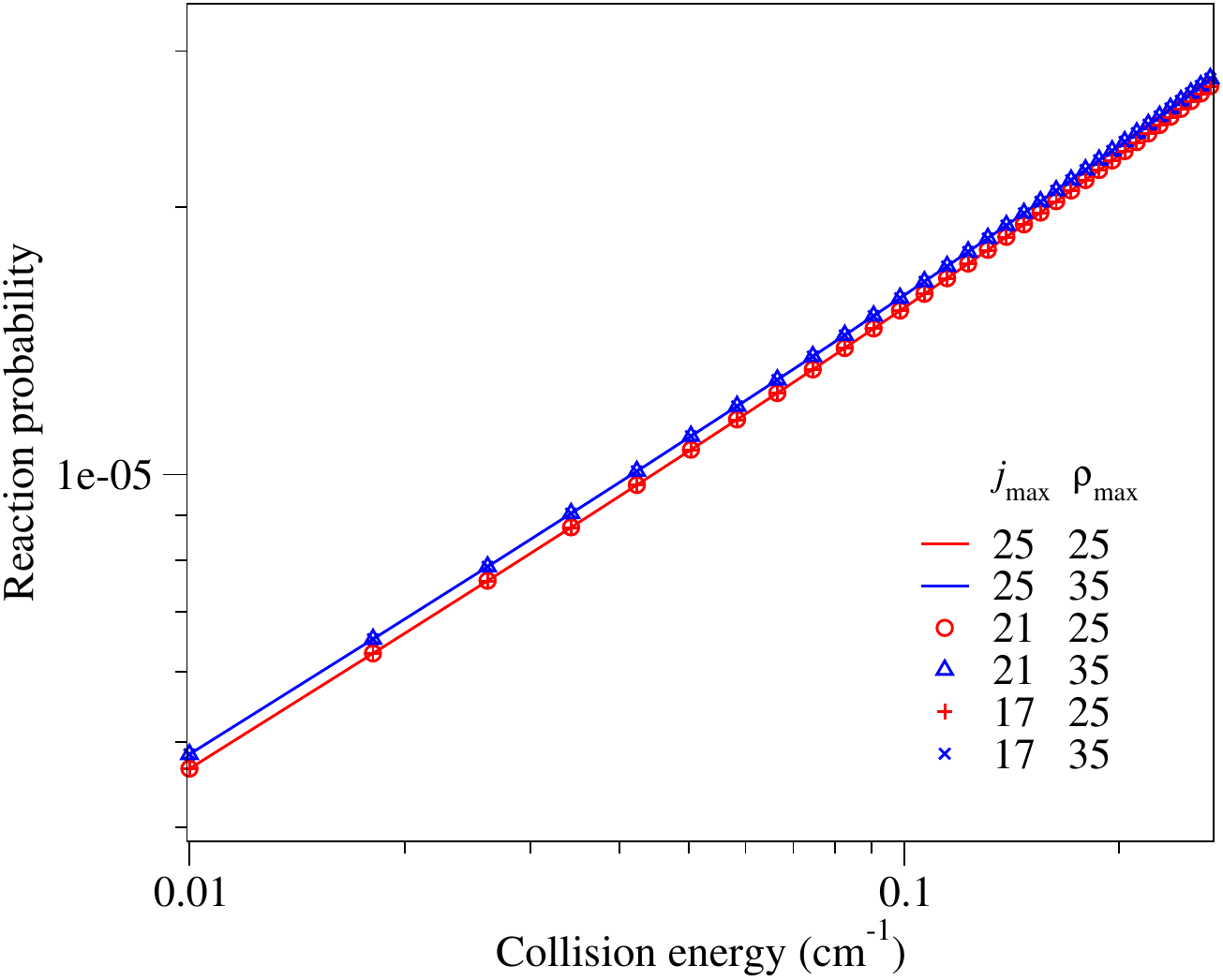}
	\renewcommand{\figurename}{Fig.}
	\caption{$J=0$ reaction probabilities for the LiF$(v=1,j=0)$ + H $\to$ Li + HF chemical reaction in the absence of external field plotted as a function of collision energy for different values of the convergence parameters $\rho_\text{max}$ and $j_\text{max}$. A total of six data sets are plotted, with $j_\text{max}=17$, 21, and 25, and $\rho_{max}=25$ $a_0$ and 35$a_0$, respectively as indicated in the legend. The basis set cutoff parameter $E_\text{max}=1.4$ eV for all calculations (increasing $E_\text{max}$ to 1.6 eV does not have a significant effect on the reaction cross sections).  The number of hyperradial propagation steps $n_\rho= 2500$ for $\rho_\text{max}=25$ $a_0$ and 3500 for $\rho_\text{max}=35$ $a_0$.}\label{fig:1}
\end{figure}

The key convergence parameters that control the accuracy of reactive scattering calculations are $E_\text{max}$ (the cutoff energy of the FD rovibrational basis set), $j_\text{max}$ (the maximum number of rotational states included in the basis),  $k_\text{max}$ -- the maximum number of BF projections of $J$ in the basis set, $\rho_\text{max}$ -- the maximum  propagation distance, and $n_\rho$ -- the number of hyperradial propagation sectors.   At low collision energies we use a complete helicity basis set, setting $k_\text{max}=J$ for any given $J$-block.  The remaining parameters were optimized following previous theoretical work on the LiF + H $\to$ Li + HF reaction at zero electric field \cite{Bala05}.

 Figure 6 shows the $J=0$ reaction probabilities for LiF + H $\to$ Li + HF as functions of collision energy computed using  $j_\text{max}=17$, 21, and 25, and $\rho_\text{max}=25$ and 35 $a_0$. We observe that truncating the rotational basis set from $j_\text{max}=25$ (the value recommended in Ref. \cite{Bala05}) to $j_\text{max}=17$ has a negligible  effect of less than 1\% on the reaction probabilities. Changing $\rho_\text{max}$ from 35~$a_0$ to 25~$a_0$ leads to a 4\% decrease in the reaction probability at $E_C=0.01$ cm$^{-1}$ and smaller changes at higher collision energies.    Based on these tests,  the following values of the convergence parameters can be used without significant loss of accuracy:  $j_\text{max}=17$,  $\rho_\text{max}=25$~$a_0$, and $n_\text{rho}=2500$.

 \begin{table}[b]
\caption{Cross sections for the LiF + H $\to$ Li + HF chemical reaction (in $a_0^2$) vs. electric field (in kV/cm) for the different basis set cutoff parameters $E_\text{max}$ (in eV). The other convergence parameters are fixed at $J_\text{max}=3$, $\rho_\text{max}=25$ $a_0$, and $j_\text{max}=17$. The number of channels $N=3828$ for $E_\text{max}=1.4$ eV and $N=4476$ for   $E_\text{max}=1.6$ eV.} 
\centering \vspace{0.2cm}
\begin{tabular}{ccc}
\hline\hline
 Electric field       &            $E_\text{max}=1.4$ eV       &       $E_\text{max}=1.6$ eV      \\
\hline
10                  &          $0.699\times 10^{-1}$     &      $0.843\times 10^{-1}$       \\
45                  &          $0.123$     &      $0.137$       \\
100                 &          $0.129$     &      $0.138$    \\
130                  &         $0.132$     &      $0.119$      \\
\hline\hline
\end{tabular}
\end{table}

 Table I lists the reaction cross sections computed for two different values of the cutoff parameter $E_\text{max}$ that controls the maximum energy of rovibrational states in the FD basis set (see main text for details).  Increasing $E_\text{max}$ enhances the accuracy of the calculations at the expense of higher computational cost of solving larger systems of coupled-channel equations. The effect of increasing $E_\text{max}$ is most pronounced at small electric fields, decreasing from $\sim$17\% at $E=10$ kV/cm to $\sim$10\% at $E>100$ kV/cm. As a compromise between accuracy and computational cost, we choose to use $E_\text{max}=1.4$ eV for production runs.  Restricting the cutoff parameter $E_\text{max}$ is likely the most significant source of convergence error in   reactive scattering calculations at low electric fields.

In field-dependent quantum reactive scattering calculations using the extended ABC code,  it is  essential to explore the convergence of reaction observables with respect to the maximum number of total angular momentum states ($J_\text{max}$) included in the basis set. Figure 7 shows the variation of the  reaction cross section with $J_\text{max}$. While the reaction cross sections at low electric fields converge rapidly, it is imperative to include at least 4 total angular momentum states $J_\text{max}=3$ in the basis sets to obtain converged results at electric fields  above 100 kV/cm. In view of the enormous computational cost of $J_\text{max}=4$ calculations ($N=5850$), we used $J_\text{max}=3$ for production runs. Thus, the biggest source of uncertainty (up to 20\% at $E=130$ kV/cm) in our results at high electric fields   is  the limited number of total angular momentum states in the basis set.

 \begin{figure}[t!]
	\centering
	\includegraphics[width=0.6\textwidth, trim = 90 0 0 0]{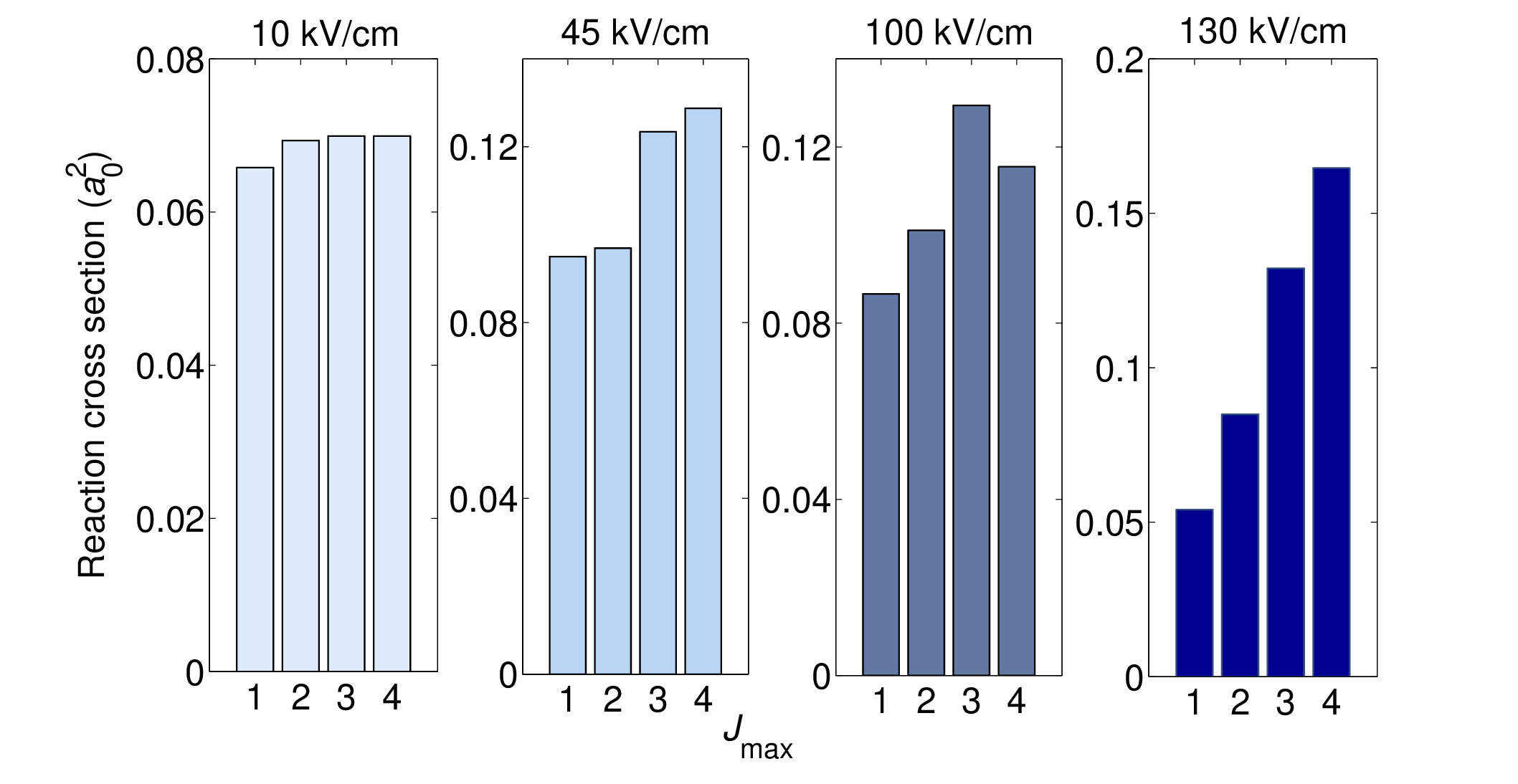}
	\renewcommand{\figurename}{Fig.}
	\caption{Convergence of reaction cross sections for LiF($v=1,j=0$) + H $\to$ Li + HF as a function of $J_\text{max}$. From left to right: $E=10$, 45, 100, and 130 kV/cm. The collision energy is 0.01 cm$^{-1}$. The other convergence parameters are: $j_\text{max}=17$,  $\rho_\text{max}=25$~$a_0$, and $n_{\rho}=2500$. }\label{fig:1}
\end{figure}

All the calculations above were performed for a fixed value of the total angular momentum projection $M=0$, which provides the dominant contribution to the total reaction cross section in the $s$-wave regime for the reactant molecules in the ground rotational state.  Test calculations performed for the electric field values of 4 and 22 kV/cm indicate that the $M=1$ contribution to the total reaction cross section amounts to a small fraction of the $M=0$ contribution (with the ratio $M=1$ to $M=0$ cross sections not exceeding  8\%), and hence can be neglected.




\begin{thebibliography}{99}
\bibitem{note1} The F + LiH channel lies 3 eV above the reactants' threshold LiF($v=1,j=0$)  and can thus be safely excluded from low-energy scattering calculations \cite{Bala05}.
\end{thebibliography}
\end{document}